\documentclass[12pt,english]{extarticle}
\usepackage{mathptmx}

\usepackage[LGR,T1]{fontenc}
\usepackage[latin9]{inputenc}
\usepackage{geometry}
\usepackage{array}
\usepackage{multirow}
\usepackage{graphicx}
\usepackage{setspace}
\usepackage{subscript}
\graphicspath{{./}}
\makeatletter

\DeclareRobustCommand{\greektext}{%
  \fontencoding{LGR}\selectfont\def\encodingdefault{LGR}}
\DeclareRobustCommand{\textgreek}[1]{\leavevmode{%
  \IfFileExists{grtm10.tfm}{}{\fontfamily{cmr}}\greektext #1}}
\DeclareFontEncoding{LGR}{}{}
\DeclareTextSymbol{\~}{LGR}{126}
\providecommand{\tabularnewline}{\\}

\usepackage{titlesec}

\titleformat{\section}[display]
  {\normalfont\scshape}
  {}
  {14pt}
  {\bf\large}
    
\titlespacing*{\section}
  {0pt}
  {0pt}
  {0pt}

\usepackage{caption}
\makeatother

\usepackage{babel}
\begin{document}
\begin{singlespace}
\section{A Comparison of Storage Ring Modeling with COSY INFINITY, ZGOUBI,
and MAD8 }

\vspace{-0pt}
Robert Hipple and Martin Berz

\end{singlespace}

Department of Physics and Astronomy, Michigan State University,
East Lansing, MI, USA \parskip=0pt

\parskip=12pt

Currently there is significant interest in the use of storage rings
to search for an electric dipole moment (EDM) in hadrons {[}1{]}.
This requires utilizing the storage ring as a precision measuring
device {[}2{]}. Part of understanding the detailed behavior of storage
rings comes from careful analysis of fringe fields {[}3{]}, but the
various tracking codes available differ in their ability to model
such behavior. It is the purpose of this paper to investigate these
differences.

A major storage ring facility actively engaged in the search for hadron
EDMs is the COoler SYnchrotron (COSY) at Forschungszentrum J\"ulich {[}4{]}.
We modeled a simplified version of this storage ring using three well-known
simulation codes \textendash{} MAD8 {[}5{]}, ZGOUBI {[}6{]} and COSY
INFINITY {[}7{]}. MAD8 is a \textquotedblleft transfer map\textquotedblright{}
code of order 2, which means that the state of the particle in phase
space is maintained as a vector, and the differential equations governing
the motion of particle through the storage ring elements are represented
by transfer maps. To track a particle through a system, one merely
needs to perform map composition. MAD8 also has the capability to
track particles symplectically using generating functions of third
order {[}8{]}.

ZGOUBI does not use the transfer map technique, but rather integrates
the Lorentz equation by time stepping based on a Taylor series in
path length. The coefficients of the Taylor series in time
are determined by an additional Taylor expansion of the magnetic field,
to fifth order maximum, if the fields are given analytically, and
by an out of plane expansion based on numerical differentiation otherwise.
ZGOUBI has few programming capabilities beyond a simple looping mechanism
to provide multiple passes through an optical system. ZGOUBI provides
support for fringe fields via Enge coefficients {[}9{]}. The software
distribution also includes a powerful post processing module called
ZPOP for plotting and data visualization.

COSY INFINITY is a combination of the advantages of the transfer map
approach and integration codes. It is primarily a transfer map code,
but utilizes integration internally to create highly accurate maps
for fringe fields {[}10{]}. Built into COSY INFINITY is an interpreter
for the specialized COSYScript programming language {[}11{]}, which
allows the researcher to simulate charged particle optics systems
to a high degree of accuracy using the techniques of Differential
Algebra. Fringe fields are specified by Enge coefficients which can
be input by the user to model actual field measurements, or a default
set of typical values can be chosen.
\begin{table}[!t]
\noindent \begin{raggedright}
\hspace*{37mm}%
\begin{tabular*}{10cm}{@{\extracolsep{\fill}}|lccc|}
\hline 
-0.9774877 & -1.078548 & 0.00 & 0.00\tabularnewline
0.03521565 & -0.9841743 & 0.00 & 0.00\tabularnewline
0.00 & 0.00 & -0.5176308 & -10.90340\tabularnewline
0.00 & 0.00 & 0.06520659 & -0.5583641\tabularnewline
\hline 
\end{tabular*}\textbf{ \ COSY INFINITY}
\par\end{raggedright}

\bigskip{}

\noindent \begin{raggedright}
\hspace*{37mm}%
\begin{tabular*}{10cm}{@{\extracolsep{\fill}}|cccc|}
\hline 
-0.9774876 & -1.0785556 & 0.00 & 0.00\tabularnewline
0.03521571 & -0.984175 & 0.00 & 0.00\tabularnewline
0.00 & 0.00 & -0.5176308 & -10.90340\tabularnewline
0.00 & 0.00 & 0.06520659 & -0.5583641\tabularnewline
\hline 
\end{tabular*}\textbf{ \ MAD8}
\par\end{raggedright}

\noindent \begin{raggedright}
\bigskip{}

\par\end{raggedright}

\noindent \raggedright{}\hspace*{38mm}%
\begin{tabular*}{10cm}{@{\extracolsep{\fill}}|cccc|}
\hline 
-0.9774910 & -1.0783900 & 0.00 & 0.00\tabularnewline
0.03521423 & -0.984180 & 0.00 & 0.00\tabularnewline
0.00 & 0.00 & -0.5176260 & -10.90340\tabularnewline
0.00 & 0.00 & .06520679 & -0.5583590\tabularnewline
\hline 
\end{tabular*}\textbf{ \ ZGOUBI}\protect\caption{First order transfer matrices for the three codes without fringe fields.
Note that COSY INFINITY and MAD8 use transfer matrices and thus naturally
agree to high accuracy, whereas ZGOUBI calculates the transfer map
as a result of integration of nearby orbits which is slightly less
accurate.\label{tab:matrices} \vspace{-10pt}}
\end{table}

\begin{figure}[!b]
\noindent \centering{}\includegraphics[scale=0.65]{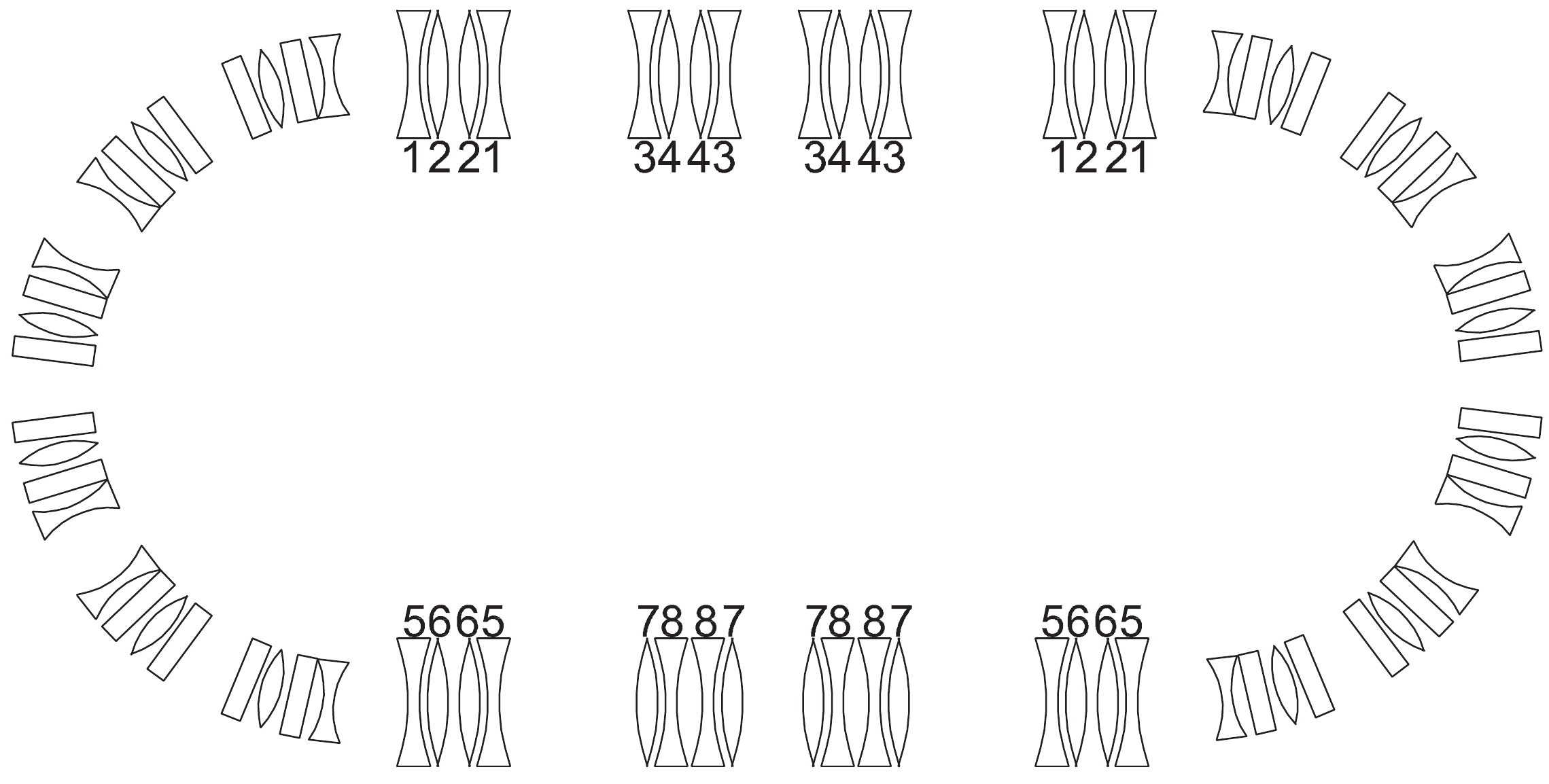}\protect\caption{Simplified model of the COSY storage ring at Forschungszentrum J\"ulich.\hspace{3.5cm}\label{fig:COSYring}}
\end{figure}

To establish a baseline, we begin with a simplified hard-edge model
of the COSY storage ring (Figure \ref{fig:COSYring}). The ring is
highly symmetric, incorporating two 40 m telescope regions, and
two 52 m arcs. Each arc is composed of three identical bending
segments, each with mirror symmetry. The bending elements are rectangular
dipoles. There are 16 sextupoles (not indicated on the figure) at
various locations around the lattice. Our model incorporates only
the bending and focusing elements -- the sextupoles in the actual lattice
are not modeled. After implementing the storage ring elements into
the three codes, we confirm that the first order transfer matrices
are essentially identical (Table \ref{tab:matrices}). 

\begin{figure}[t]
\vspace{-2.3cm}
\noindent \raggedright{}\includegraphics[scale=0.24]{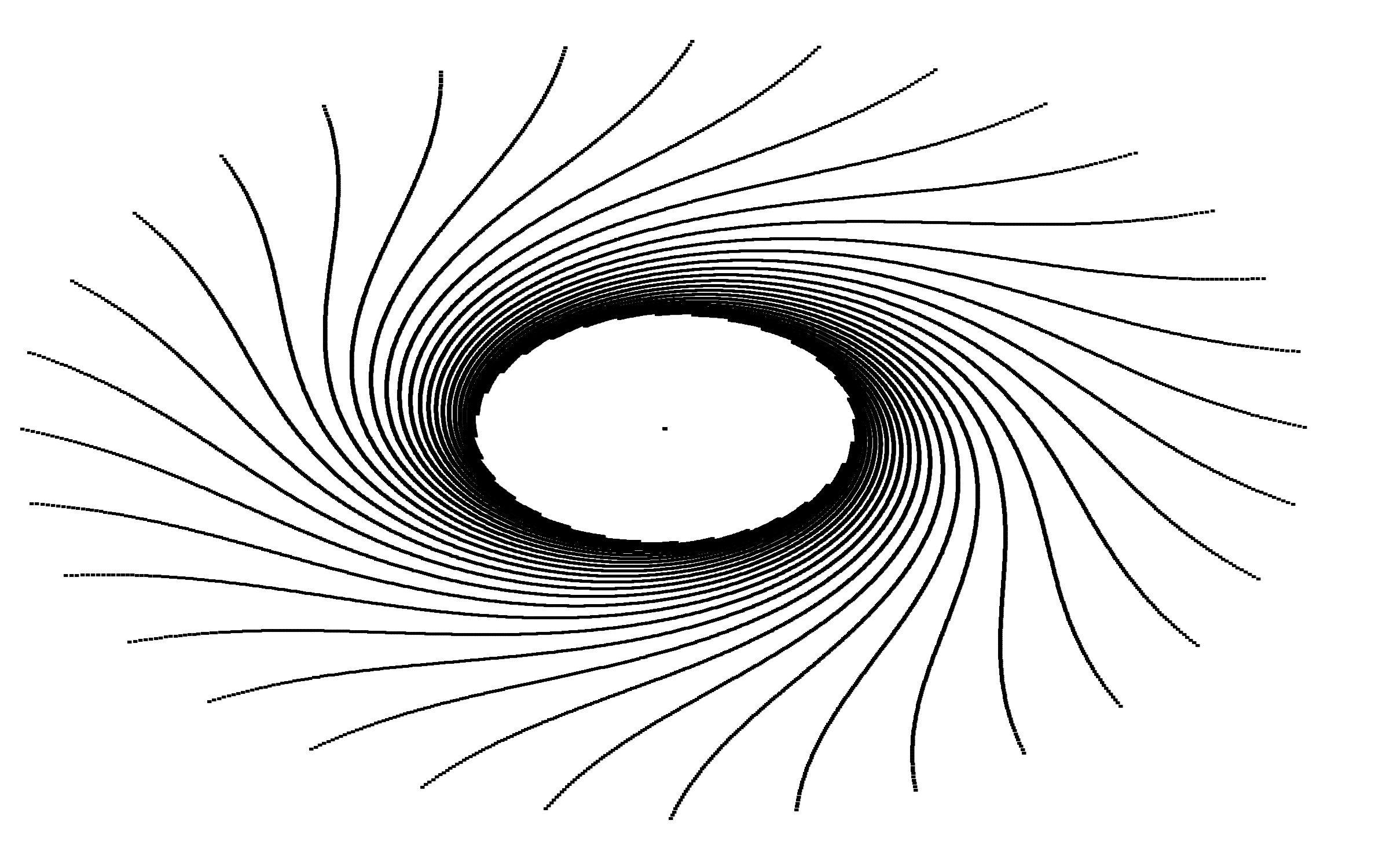}\includegraphics[scale=0.22]{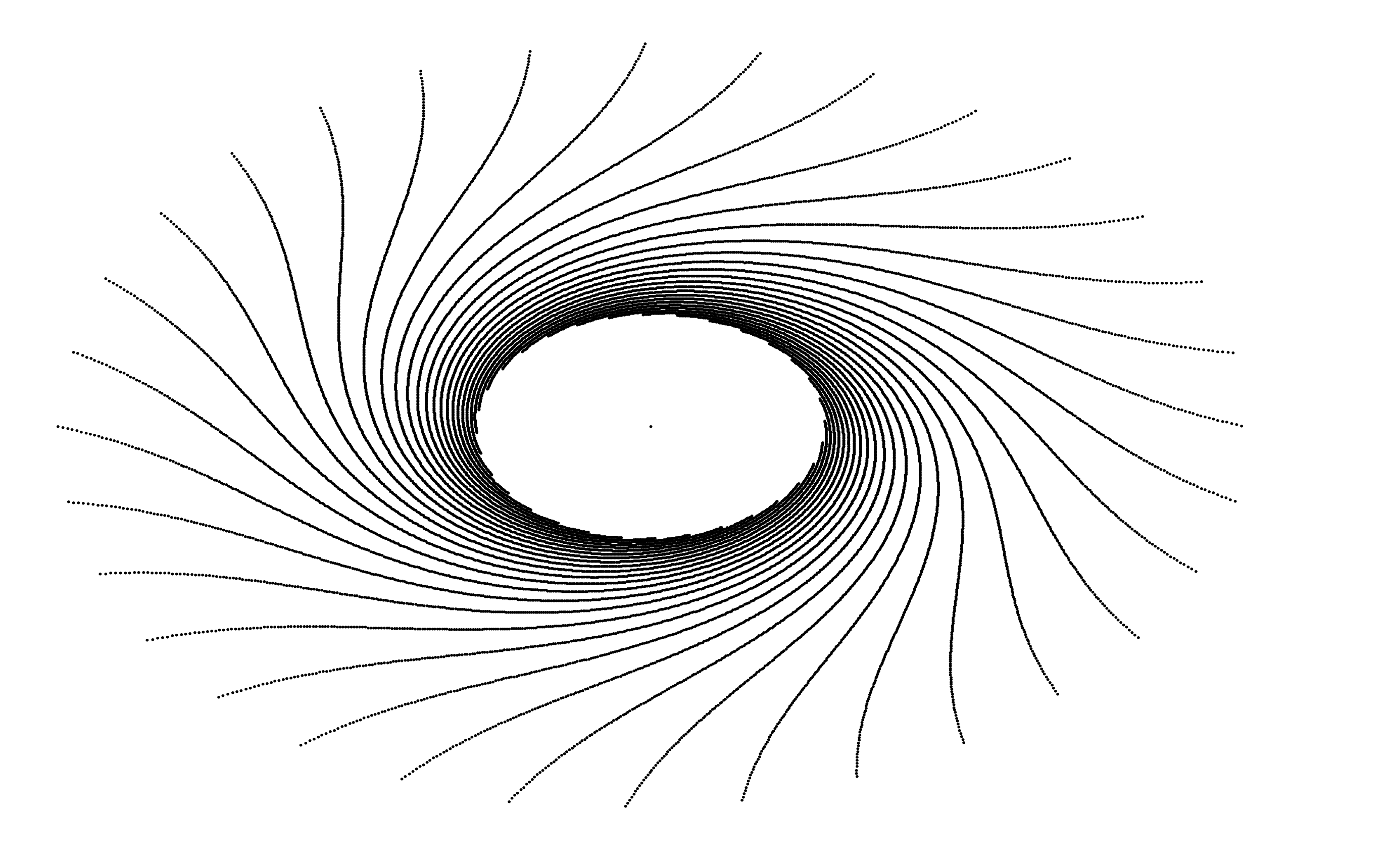}\includegraphics[bb=67bp 50bp 540bp 735bp,scale=0.25]{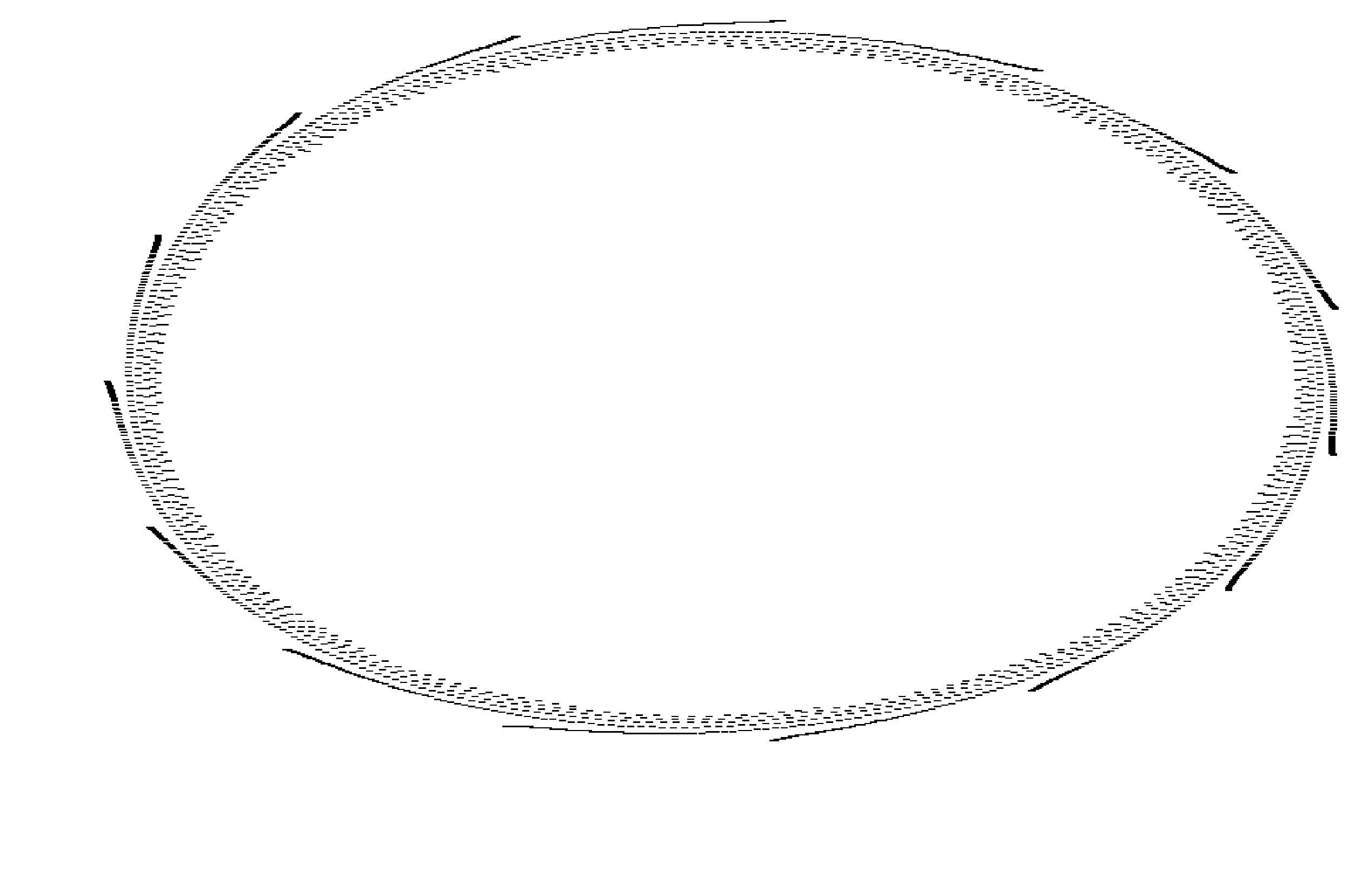}\protect\caption{Second order tracking in the horizontal plane created by COSY INFINITY
(left) and MAD8 (center), compared with ZGOUBI (right) which does
not use transfer maps, under identical initial conditions and lattice
parameters. No symplectification is used, showing that second order
is insufficient to describe the dynamics.\label{fig:nosymp} }
\end{figure}

Having verified that our lattices agree to first order, we perform
some initial tracking runs and compare the results. Figure \ref{fig:nosymp}
shows the tracking pictures for a single 970 MeV/c proton over 2000
turns with cosine-like initial conditions at transverse amplitude
< 1 cm. There is quite good agreement between MAD8 and COSY INFINITY
when performing tracking at MAD8's maximum order of 2, including the
inward spiral characteristic of non-symplectic tracking. The output
for ZGOUBI also shows a small violation of symplecticity, but apparently
ZGOUBI's accuracy is higher than that of second order transfer maps.

\begin{figure}[!t]
\noindent \centering{}\includegraphics[scale=0.25]{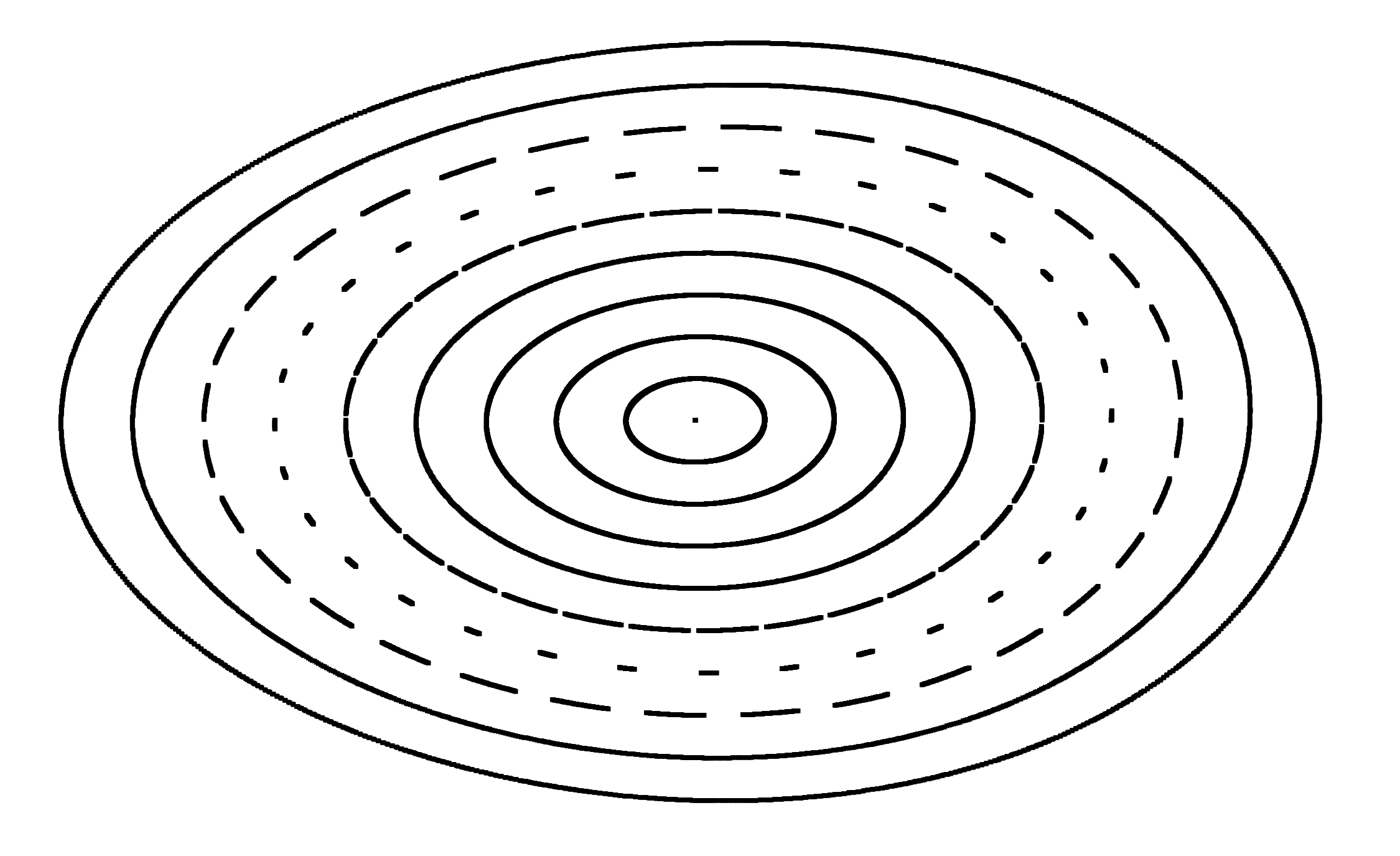}\hspace*{4mm}\includegraphics[scale=0.25]{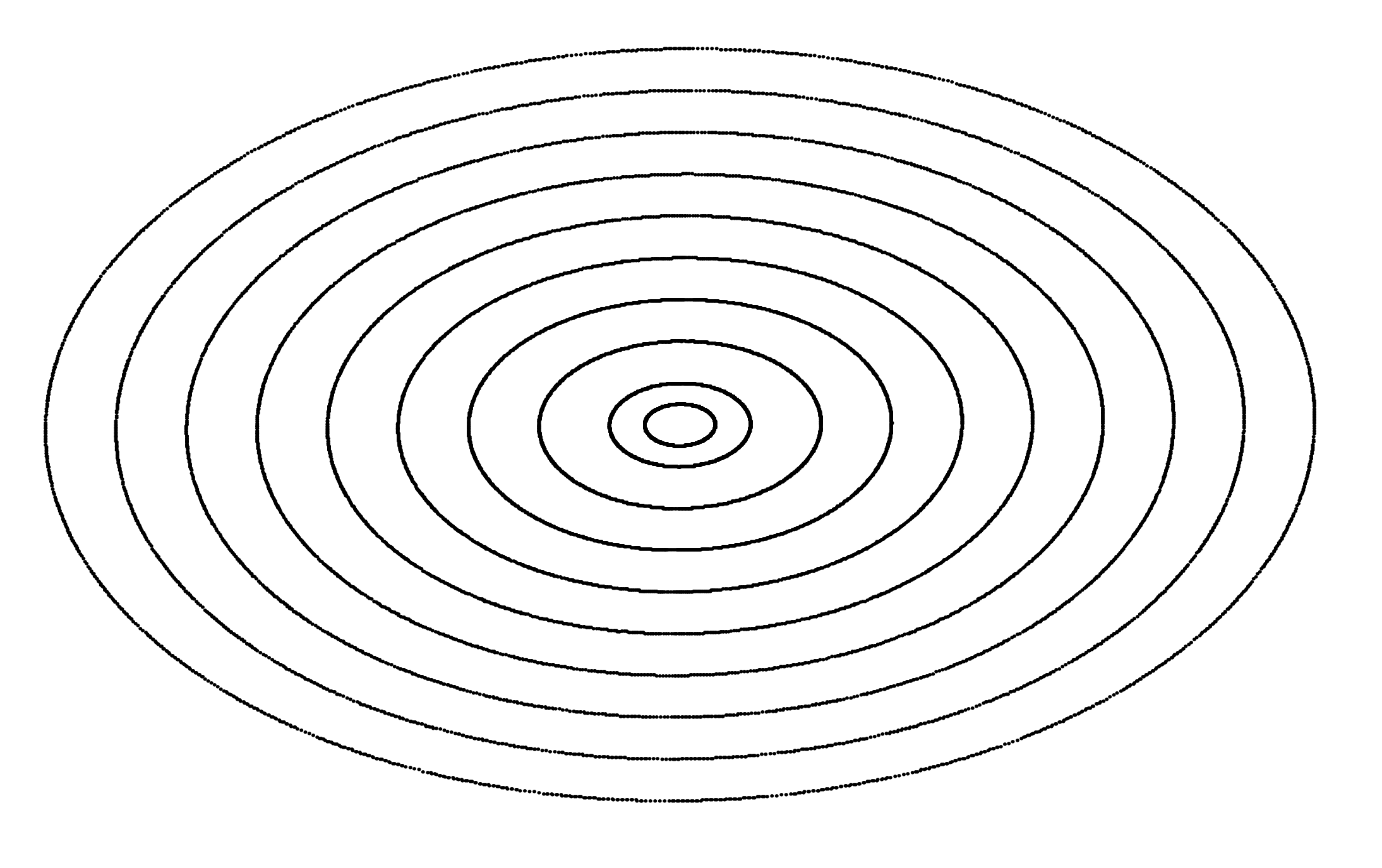}\hspace*{4mm}\includegraphics[scale=0.25]{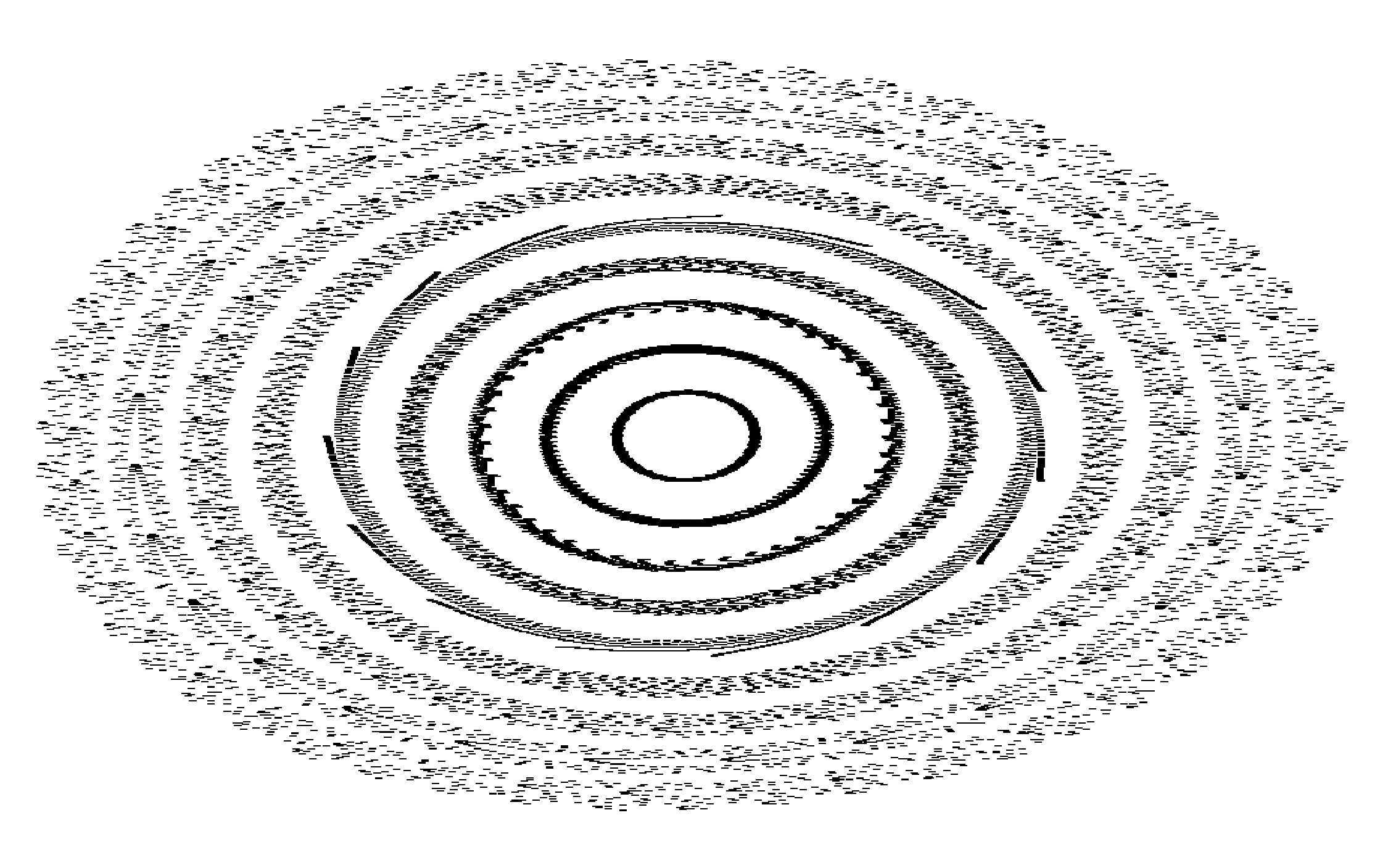}\protect\caption{Varying transverse initial conditions from 1 cm to 9 cm in the horizontal
plane for second order tracking with COSY INFINITY (left) and MAD8
(center) and compared with ZGOUBI (right). Symplectic tracking is
enabled in COSY INFINITY and MAD8. This is not available in ZGOUBI,
which shows a widening of the orbit bands indicative of violation
of symplecticity.\label{fig:symp}}
\end{figure}
\begin{figure}[!t]
\noindent \centering{}\includegraphics[scale=0.25]{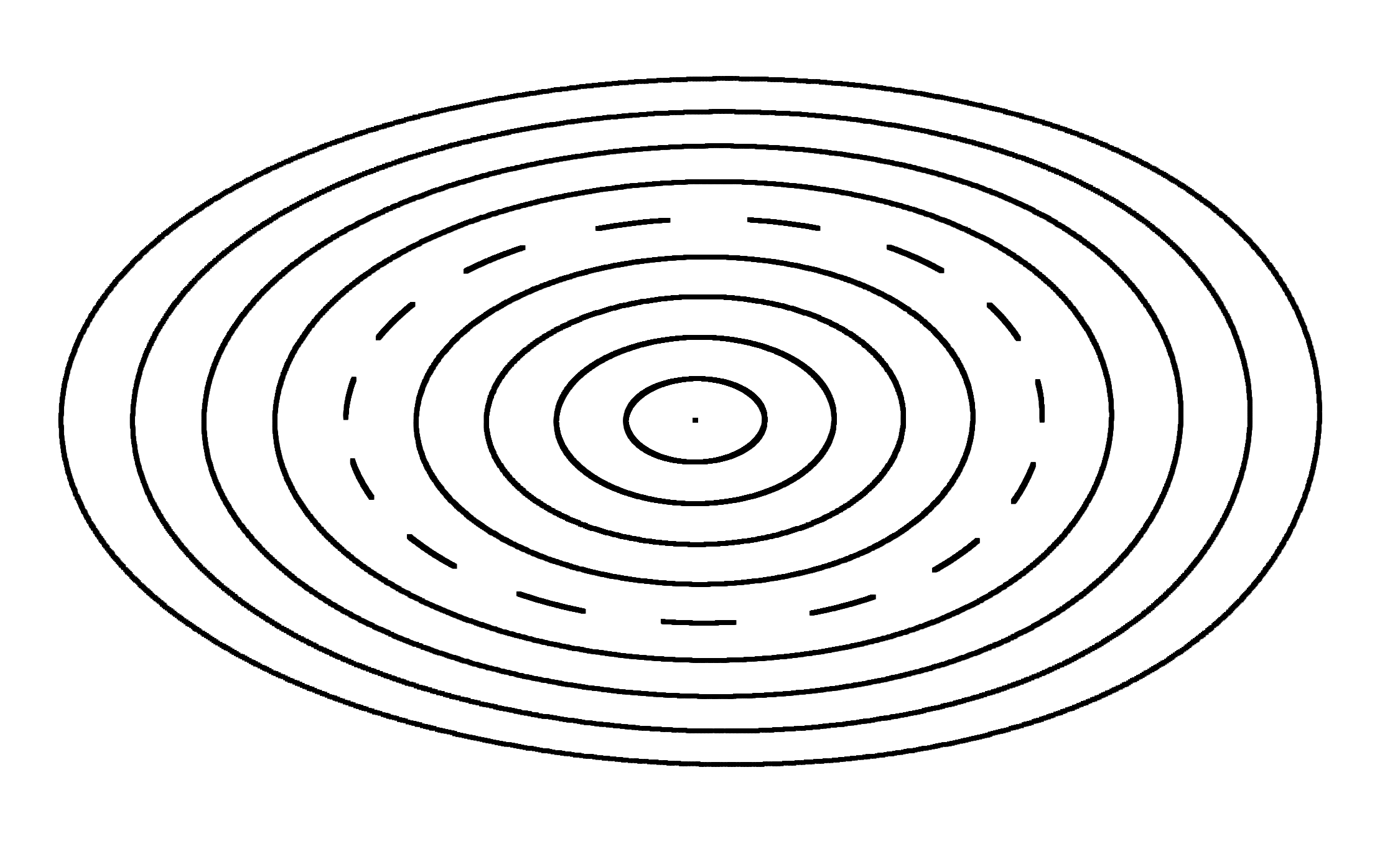}\protect\caption{Same tracking as Figure \ref{fig:symp}, now utilizing COSY INFINITY\textquoteright s
9th order tracking without symplectification, still without fringe
fields. The high order of the map and resulting accuracy avoids the
violation of symplecticity visible in Figure 2. Even to this high
order, there is very little nonlinearity evident in the dynamics.\label{fig:cosy9th} \vspace{-10pt}}
\end{figure}

With COSY INFINITY and MAD8 we can enable symplectic tracking {[}12{]},
which ZGOUBI does not support. Figure \ref{fig:symp} exhibits tracking
pictures with symplectic tracking enabled. They are virtually identical
for COSY INFINITY and MAD8, the phase space orbits having a well-defined
elliptical shape. This establishes a baseline between COSY INFINITY,
ZGOUBI and MAD8. Turning to the question of dynamic aperture, we track
particle orbits in 1 cm steps from the reference orbit out to 9 cm \textendash{}
twice the physical aperture of the actual ring. We see little significant
deviation from linear behavior across all three codes. So far, with
COSY INFINITY and MAD8, we are still tracking with second order transfer
maps. With COSY INFINITY we can push to higher orders with very little
extra processing time. Figure \ref{fig:cosy9th} is the same tracking
run, this time to 9th order in the transfer maps. Still there is very
little nonlinearity evident in the dynamics, and the dynamic aperture
appears unlimited. 

Judging from these results alone, one might conclude that the lattice
is remarkably stable. We now investigate the effect of incorporating
fringe fields. COSY INFINITY has the convenience of enabling a default
set of fringe field profiles with the single command ``FR 3''.
Table \ref{tab:matricesFF} shows the COSY INFINITY first order transfer
map with fringe fields turned on. Notice that the motion in the horizontal
direction is now \textit{unstable} (|Trace| > 2). 
\begin{table}[t]
\hspace{0.2\textwidth}%
\begin{tabular*}{10cm}{@{\extracolsep{\fill}}|cccc|}
\hline 
\textbf{-0.9739104} & 1.954368 & 0.00 & 0.00\tabularnewline
0.01832738 & \textbf{-1.063567} & 0.00 & 0.00\tabularnewline
0.00 & 0.00 & -0.7993542 & -6.705731\tabularnewline
0.00 & 0.00 & 0.05219644 & -0.8131372\tabularnewline
\hline 
\end{tabular*}\textbf{ \ COSY INFINITY}

\bigskip

\hspace{0.2\textwidth}%
\begin{tabular*}{10cm}{@{\extracolsep{\fill}}|cccc|}
\hline 
\textbf{-0.9618310} & 2.08049 & 0.00 & 0.00\tabularnewline
.02259699 & \textbf{-1.089850} & 0.00 & 0.00\tabularnewline
0.00 & 0.00 & -0.209283 & 17.4232\tabularnewline
0.00 & 0.00 & -.05576599 & -.13560600\tabularnewline
\hline 
\end{tabular*}\textbf{ \ ZGOUBI}\protect\caption{First-order transfer matrices for COSY INFINITY and ZGOUBI with full
fringe fields enabled. The horizontal trace (the sum of the bold matrix
elements) is greater than 2 in magnitude, which indicates instability.\label{tab:matricesFF}\vspace{-10pt}}
\end{table}

We can confirm this by adding identical fringe fields to ZGOUBI and
checking the computed transfer matrix (Table \ref{tab:matricesFF}).
A similar horizontal instability is indicated in the ZGOUBI transfer
matrix as well. It is apparent that some adjustment of the magnet
strengths is necessary to stabilize the beam. Without making any specific
assumptions regarding which adjustments to make, we choose an arbitrary,
minimally invasive approach which respects the existing symmetries
of the lattice. The numbered elements in Figure \ref{fig:COSYring}
indicate quadrupoles which are equal in field strength. By preserving
this equality, we maintain the symmetry that was designed into the
lattice. This allows 8 degrees of freedom. Since we are trying to
match 8 elements of the transfer matrix, we require only six degrees
of freedom due to the unity determinant. By utilizing an implementation
of the MINPACK LMDIF optimizer built into COSY INFINITY {[}13{]},
we find a (not necessarily unique) set of quadrupole strength multipliers
which achieve our objective:

\vspace{-12pt}

\begin{center}
\begin{tabular}{cccc}
\textgreek{l}\textsubscript{1}= 1.022183, & \textgreek{l}\textsubscript{2}= 1.019331, & \textgreek{l}\textsubscript{3}= 1.009453, & \textgreek{l}\textsubscript{4}=1.019948,\tabularnewline
\textgreek{l}\textsubscript{5}= 0.999410, & \textgreek{l}\textsubscript{6}= 1.019427, & \textgreek{l}\textsubscript{7}= 1.031603, & \textgreek{l}\textsubscript{8}=1.013424.\tabularnewline
\end{tabular}
\end{center}

\vspace{-12pt}

We see this requires an adjustment of only a few percent in the quadrupole
strengths. Implementing these changes into all three codes, we successfully
restore the lattice to its original design transfer matrix. 

\begin{figure}[t]
\begin{spacing}{0}
\noindent \centering{}\includegraphics[scale=0.3]{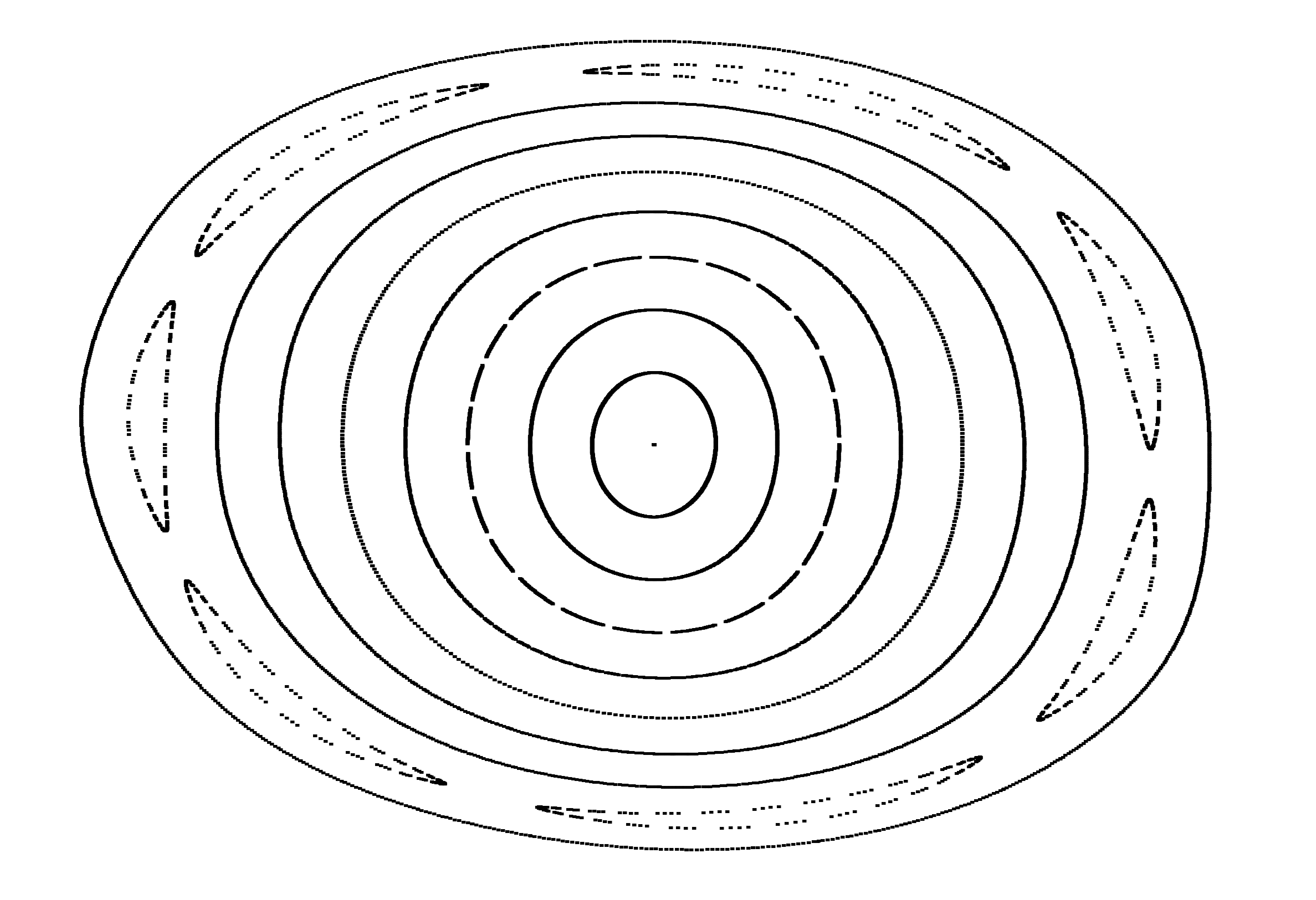}\hspace*{10mm}\includegraphics[scale=0.3]{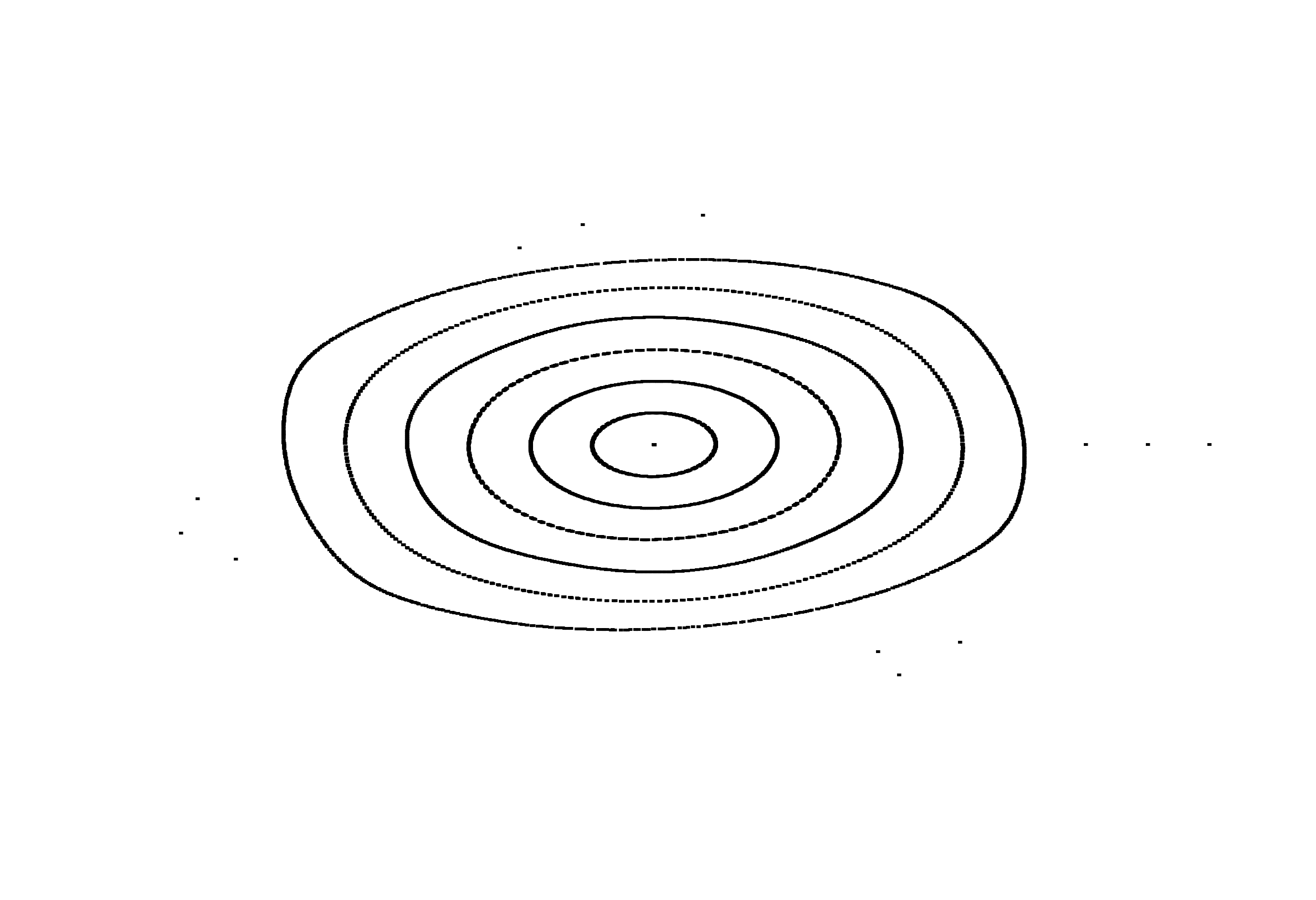}\protect\caption{COSY INFINITY 9th order tracking with full fringe field simulation
capabilities in vertical (left) and horizontal (right) planes. \label{fig:cosyFF}}
\end{spacing}
\end{figure}

Figure \ref{fig:cosyFF} shows the resulting tracking pictures for
COSY INFINITY. This run was 9th order in the transfer matrices, with
the fringe fields enabled for all lattice elements. In the horizontal
direction, nonlinearities are readily apparent. The shape of the phase
space ellipses are seen to vary with increasing distance from the
reference orbit. At 8 cm, seven \textquotedblleft islands\textquotedblright{}
of stability appear. In the vertical direction, there is substantial
beam loss at distances > 6 cm. Neither of these phenomena are visible
in the hard-edge approximation.
\begin{figure}[!t]
\begin{singlespace}
\noindent \centering{}\includegraphics[scale=0.44]{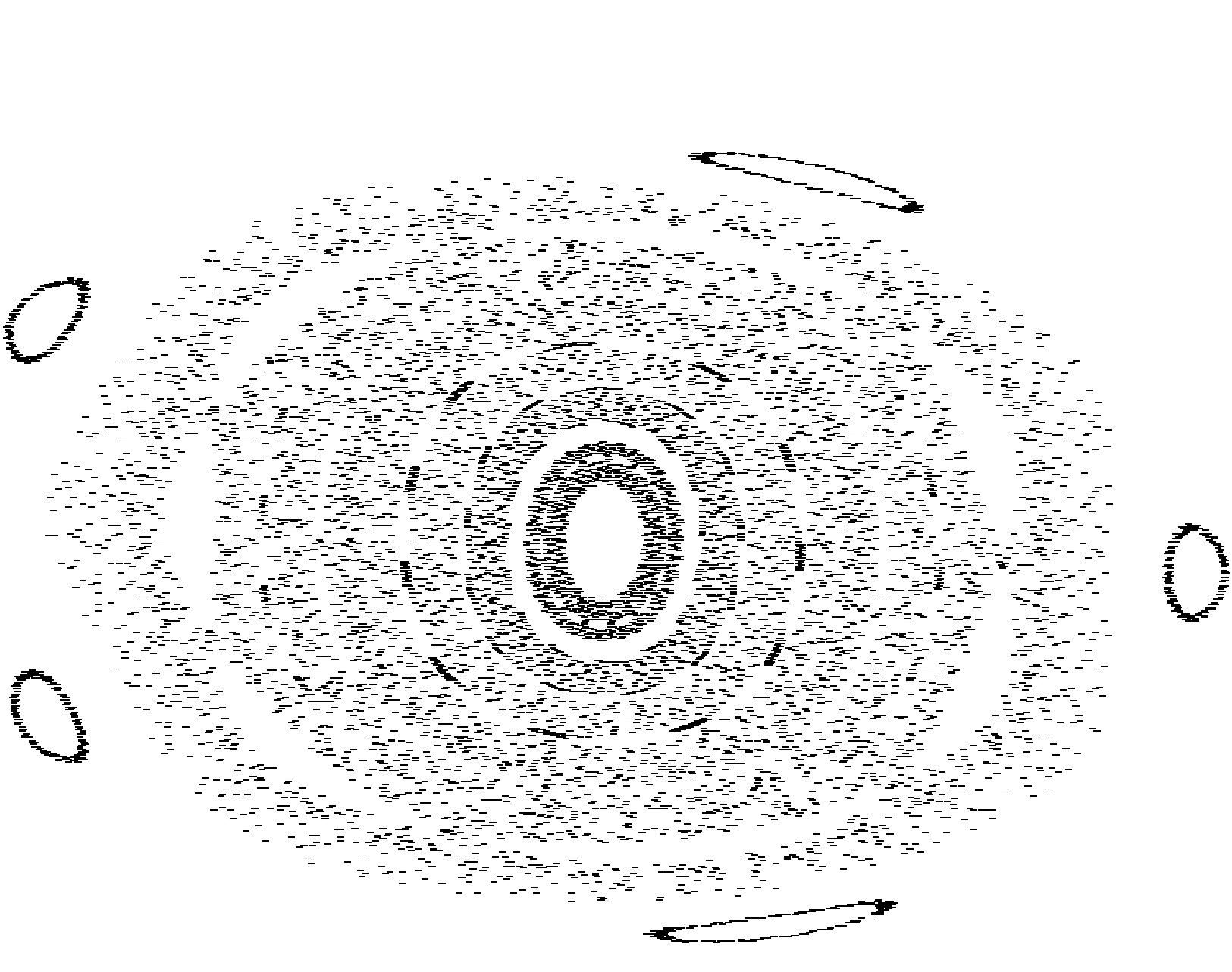}\hspace*{10mm}\includegraphics[scale=0.44]{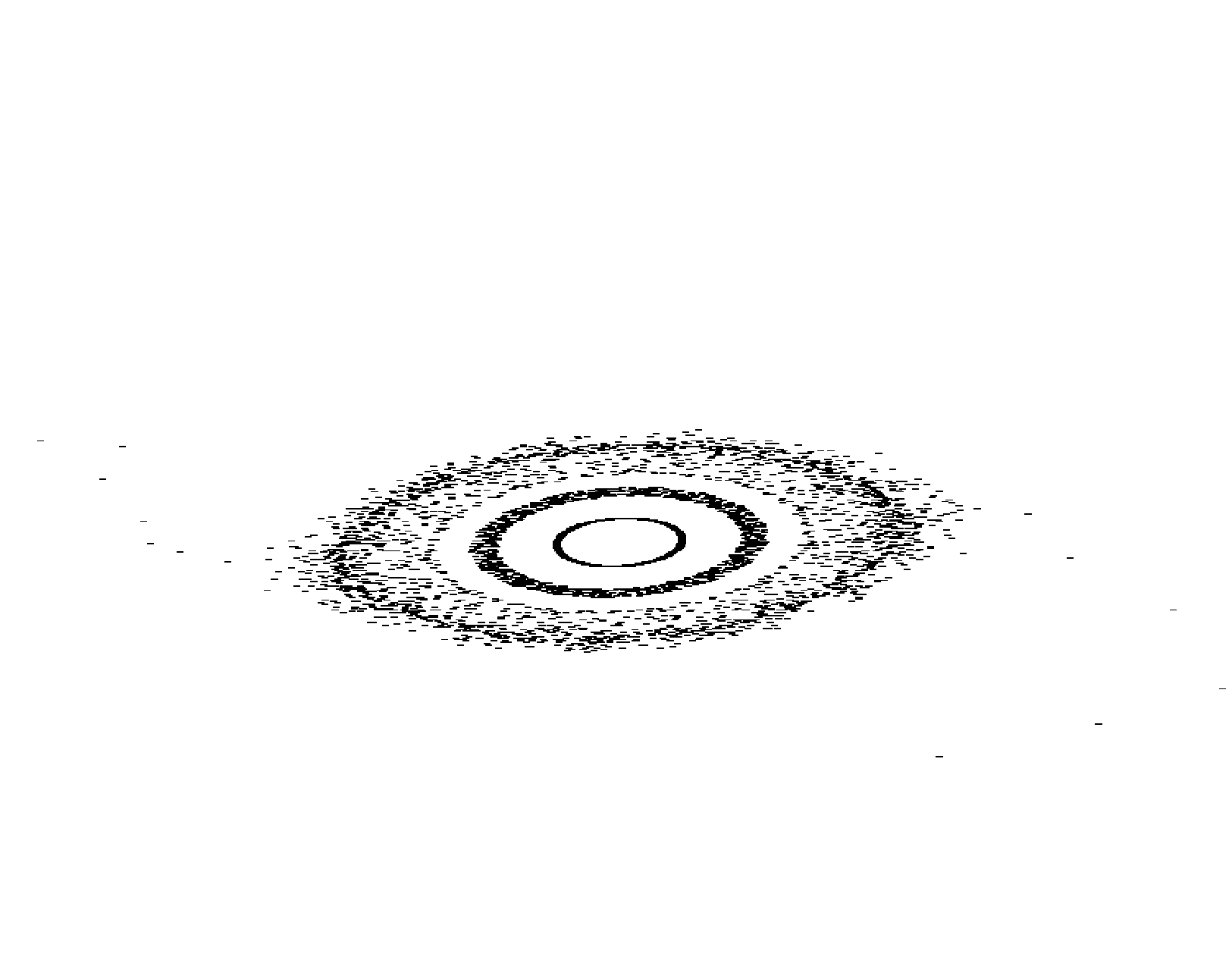}\protect\caption{ZGOUBI tracking with full fringe field simulation capabilities in
vertical (left) and horizontal (right) planes. The fringe field Enge
coefficients are identical to those used in COSY INFINITY. The rough
structure and stability boundaries are similar to those of Figure
5, but symplecticity violations are apparent. \label{fig:zgoubiFF}}
\end{singlespace}
\vspace{-10pt}
\end{figure}

Figure \ref{fig:zgoubiFF} shows the same tracking experiment using
ZGOUBI. It is interesting to note that the seven \textquotedblleft islands\textquotedblright{}
revealed by COSY INFINITY are visible with ZGOUBI as well but these
have the appearance of \textquotedblleft gaps\textquotedblright{}
in the tracking picture. In the vertical direction, ZGOUBI also agrees
with COSY INFINITY\textquoteright s prediction of substantial beam
loss at distances greater than 6 cm.

In conclusion, modeling a storage ring without taking fringe fields
into account provides an overly optimistic dynamic aperture and an
incomplete picture of the dynamics. Although COSY INFINITY and ZGOUBI
differ in their methods of tracking, both codes hint at similar nonlinear
dynamics and agree very well for the linear motion with fringe fields.
This agreement across codes, despite completely different tracking
methods, gives confidence that the predictions made reflect real physics.
There is a substantial difference in run time, however (Table \ref{tab:CPU}),
and for moderately large numbers of turns and particles, COSY INFINITY
is about three orders of magnitude faster than the identical simulation
in ZGOUBI. MAD8, which does a good job of modeling within the SCOFF
(Sharp Cutoff Fringing Field) approximation, has limited fringe field
capabilities. This limits its usefulness for lower momentum tracking
over many turns, precisely the type of tracking required for mid-size
storage rings {[}14{]}.\\ 

\begin{table}[t]
\noindent \centering{}%
\begin{tabular}{|c|c|c|}
\hline 
\multirow{2}{*}{Number of Turns} & \multicolumn{2}{c|}{CPU time (seconds) }\tabularnewline
\cline{2-3} 
 & COSY INFINITY & ZGOUBI\tabularnewline
\hline 
$10^{3}$ & 25.082 & 183.78\tabularnewline
\hline 
$10^{4}$ & 25.297 & 1831.4\tabularnewline
\hline 
$10^{5}$ & 27.132 & 18717\tabularnewline
\hline 
$10^{6}$ & 45.343 & \tabularnewline
\hline 
$10^{7}$ & 228.26 & \tabularnewline
\hline 
$10^{8}$ & 2049.4 & \tabularnewline
\hline 
$10^{9}$ & 20193 & \tabularnewline
\hline 
\end{tabular}\protect\caption{Comparison of tracking execution times of COSY INFINITY and ZGOUBI
at their respective maximum precisions (order 9 transfer map for COSY
INFINITY, 5th order Taylor series integration for ZGOUBI). ZGOUBI
execution times are proportional to the number of turns and around
0.187 seconds per turn per particle. COSY INFINITY requires an initial investment
in the computation of a transfer map, but for larger turn numbers
tracks for 1/50000 of a second per turn. \label{tab:CPU}\vspace{-10pt}}
\end{table}

References:

{[}1{]} A Lehrach \textit{et al}, XIV Advanced Research Workshop on High Energy Spin Physics (2012) p. 287.\parskip=0pt

{[}2{]} A Lehrach, private communication.

{[}3{]} H Wollnik, \textquotedblleft Optics of charged particles\textquotedblright ,
(Academic Press, Orlando, 1987).

{[}4{]} U Bechstedt \textit{et al}, Nuclear instruments and methods in physics
research B \textbf{113} (1996), p. 26.

{[}5{]} H Grote and C Iselin, \textquotedblleft The MAD program
(methodical accelerator design), version 8.13/8, user's reference
manual\textquotedblright , CERN/SL/90-13 (AP) (Rev. 4), (CERN, Geneva, 2012).

{[}6{]} F M\'{e}ot, \textquotedblleft ZGOUBI users' guide\textquotedblright, C-AD/AP/470, (Brookhaven National Laboratory, Upton, 2013).

{[}7{]} M Berz and K Makino, ``COSY INFINITY 9.1 beam
physics manual'', MSUHEP 060804-rev, (Michigan State University, East Lansing, 2013), http://cosyinfinity.org. 

{[}8{]} C Iselin, \textquotedblleft The MAD program (methodical
accelerator design), version 8.13, physical methods manual\textquotedblright ,
CERN/SL/92 (AP), (CERN, Geneva, 2012).

{[}9{]} H Enge in \textquotedblleft Focusing of charged particles\textquotedblright ,
ed. A Septier, (Academic Press , New York, 1967), p. 203.

{[}10{]} M Berz, ``Modern map methods in particle beam physics'', (Academic
Press, San Diego, 1999).

{[}11{]} M Berz and K Makino, \textquotedblleft COSY INFINITY 9.1 programmer\textquoteright s
manual\textquotedblright , MSUHEP 101214, (Michigan State University, East Lansing, 2011), http://cosyinfinity.org.

{[}12{]} S Manikonda, M Berz and B Erd\'{e}lyi,  Institute
of Physics CS \textbf{175} (2004), p. 299.

{[}13{]} JJ Mor\'{e}, BS Garbow and KE Hillstrom, \textquotedblleft User
guide for MINPACK-1\textquotedblright, ANL-80-74, (Argonne National Laboratory, Argonne, 1980).

{[}14{]} We are grateful for financial support from the US Department of Energy
under grant DE-FG02-08ER41546. We thank Kyoko Makino at Michigan State University for helpful comments and
suggestions, and Denis Zyuzin and Marcel Rosenthal at Forschungszentrum J\"ulich for
their assistance in procuring the lattice model.
\end{document}